\documentclass[aps,prb,twocolumn,groupedaddress,showpacs,amssymb,amsmath]{revtex4-1}
\usepackage{prettyref}
\usepackage{graphicx}
\usepackage{verbatim}   
\usepackage{color}      
\usepackage{subfigure}  
\usepackage{hyperref}

\begin{document}


\title{Cumulant generating function formula of heat transfer in ballistic
system with lead-lead coupling}

\author{Huanan Li}

\email{g0900726@nus.edu.sg}

\affiliation{Department of Physics and Center for Computational Science and Engineering,
National University of Singapore, Singapore 117542, Republic of Singapore }

\author{Bijay Kumar Agarwalla}

\affiliation{Department of Physics and Center for Computational Science and Engineering,
National University of Singapore, Singapore 117542, Republic of Singapore }

\author{Jian-Sheng Wang}

\affiliation{Department of Physics and Center for Computational Science and Engineering,
National University of Singapore, Singapore 117542, Republic of Singapore }

\date{24 Aug 2012}
\begin{abstract}
Based on two-time observation protocol, we consider heat transfer
in a given time interval $t_{M}$ in lead-junction-lead system taking
coupling between the leads into account. In view of the two-time observation,
consistency conditions are carefully verified in our specific family
of quantum histories. Furthermore, its implication is briefly explored.
Then using nonequilibrium Green's function (NEGF) method, we obtain
an exact formula for the cumulant generating function for heat transfer
between the two leads valid in both transient and steady-state regimes.
Also, a compact formula for the cumulant generating function in the
long-time limit is derived, for which the Gallavotti-Cohen (GC) fluctuation
symmetry is explicitly verified. In addition, a brief discussion of
Di Ventra's trick regarding whether the effect of the repartitioning
procedure of the total Hamiltonian on nonequilibium steady-state current
fluctuation exists is given. All kinds of properties of nonequilibrium
current fluctuations such as the fluctuation theorem in different
time regimes could be readily given according to these exact formulas.
\end{abstract}



\maketitle

\section{INTRODUCTION\label{sec:INTRODUCTION}}

The physics of nonequilibrium many-body systems is one of the most
rapidly expanding areas of theoretical physics. In the combined field
of non-equilibrium state and statistics, the distribution of transferred
charges in electronic case or heat in phononic case, the so-called
full counting statistics (FCS), plays an important role, according
to which we could understand the general features of currents and
their fluctuations. Also, it is well known that the noise generated
by nanodevices contains valuable information on microscopic transport
processes not available from only transient or steady current. In
FCS, the key object we need to study is the cumulant generating function
(CGF) which presents high order correlation information of the corresponding
system for the transferred quantity.

The study of the FCS started from the field of electronic transport
pioneered by Levitov and Lesovik, who presented an analytical result
for the CGF in the long-time limit \cite{Levitov1993}. After that,
many works followed in electronic FCS \cite{Belzig2001,Nazarov2003,Schonhammer2007,Esposito2009},
while much less attention is given to phonon transport. Saito and
Dhar are the first ones to borrow this concept to thermal transport
\cite{Saito2007}. Later, Ren and his co-workers gave results for
two-level systems \cite{Ren2010}. And very recently, transient behavior
and long-time limit of CGF have been obtained in lead-junction-lead
harmonic networks both classically and quantum-mechanically using
Langevin equation method and NEGF method, respectively \cite{Saito2011,Wang2011,Agarwalla2012}.
Experimentally the FCS in electronic case has been carefully studied
and the cumulants to very high order have been successfully measured
in quantum-dot systems \cite{Flindt2009,Utsumi2010}. In principle, similar measurements
could be carried out for thermal transport, e.g., in a nanoresonator
system.

Through modern nanoscale technology, small junction is easily realized
such as in certain nanoscale systems, for instance, a single molecule
or, in general, a small cluster of atoms between two bulk electrodes.
In that case, the electrode surfaces of the bulk conductors may be
separated by just a few angstroms so that some finite electronic coupling
between the two surfaces is present taking into account the long-range
interaction. In order to study steady current in this situation, Di
Ventra suggest that we can choose our ``sample'' region (junction)
to extend several atomic layers inside the bulk electrodes where screening
is essentially complete so that the above coupling could be negligible
\cite{Ventra2008}. This trick has been checked in some restricted
case in a recent work \cite{Li2012}. However, whether this intuitively
reasonable trick applies to all the higher cumulants of heat transfer
in steady state is still a question. Therefore we will briefly discuss
this question at the end of the article. And obviously this trick
can not be applied to study transient behavior of all the cumulants
of heat transfer. Thus, in this work, we want to construct the CGF
of heat transfer in a general case taking into account the coupling between the lead
explicitly, which include the information of all the cumulant of heat
transfer. Also, the transient behavior and the long-time limit of
CGF of heat transfer would be dealt with on an equal footing based
on the NEGF method.

This paper is organized as follows. We start in Sec.~\ref{sec:MODEL-AND-CONSISTENT}
by introducing the model and the consistent quantum framework. And
then we employ this framework to derive the CGF of heat transfer across
the junction in Sec.~\ref{sec:CUMULANT-GENERATING-FUNCTION}. In
addition, we obtain a compact form of the long-time limit of the CGF
in Sec.~\ref{sec:THE-STEADY-STATE-CGF}, based on which the steady-state
fluctuation theorem (SSFT) is checked and the generalized Caroli formula
closely related to the first cumulant is recovered in Sec.~\ref{sec:THE-SSFT-AND}.
At the end, we will briefly discuss Di Ventra's trick.

\section{MODEL AND CONSISTENT QUANTUM FRAMEWORK\label{sec:MODEL-AND-CONSISTENT}}

We consider the lead-junction-lead model initially prepared in product
state $\rho^{ini}=\frac{e^{-\beta_{L}H_{L}}}{\mathrm{Tr}\left(e^{-\beta_{L}H_{L}}\right)}\otimes\frac{e^{-\beta_{C}H_{C}}}{\mathrm{Tr}\left(e^{-\beta_{C}H_{C}}\right)}\otimes\frac{e^{-\beta_{R}H_{R}}}{\mathrm{Tr}\left(e^{-\beta_{R}H_{R}}\right)}$.
We can imagine that left lead $\left(L\right)$, center junction $\left(C\right)$,
and right lead $\left(R\right)$ in this model were in contact with
three different heat baths at the inverse temperatures $\beta_{L}\equiv\left(k_{B}T_{L}\right)^{-1}$,
$\beta_{C}\equiv\left(k_{B}T_{C}\right)^{-1}$ and $\beta_{R}\equiv\left(k_{B}T_{R}\right)^{-1}$
respectively, for time $t<0$. At time $t=0$, all the heat bath are
removed, and coupling of the center junction with the leads $H_{LC}\equiv u_{L}^{T}V^{LC}u_{C}$
and $H_{CR}\equiv u_{C}^{T}V^{CR}u_{R}$ and the lead-lead coupling
term $H_{LR}=u_{L}^{T}V^{LR}u_{R}$ are switched on abruptly. Now
the total Hamiltonian of the lead-junction-lead system will become
\begin{align}
H_{tot}= & H_{L}+H_{C}+H_{R}+H_{LC}+H_{CR}+H_{LR},\label{eq:total H}
\end{align}
where $H_{\alpha}=\frac{1}{2}p_{\alpha}^{T}p_{\alpha}+$$\frac{1}{2}u_{\alpha}^{T}K^{\alpha}u_{\alpha},\;\alpha=L,C,R$,
represents coupled harmonic oscillators, $u_{\alpha}\equiv\sqrt{m}x_{\alpha}$
and $p_{\alpha}$ are column vectors of transformed coordinates and
corresponding conjugate momenta in region $\alpha$. The superscript
$T$ stands for matrix transpose.

In order to extract information on heat transfer, we introduce two-time
observation protocol in the process of the evolution of the system;
that is, at time $t=0^{-}$, we carry out the measurement of energy
of the left lead associated with the operator $H_{L}$, obtaining
the result to be the eigenvalue $a$ of $H_{L}$, and measure it again
at time $t=t_{M}$, obtaining the eigenvalue $b$ of $H_{L}$. Here
the measurement is in the sense of quantum measurement of von Neumann
\cite{Neumann1955}. This quantum history \cite{Griffiths2002}, that
is a sequence of quantum events at successive times, is represented
by the product projector $Y=P_{0^{-}}^{a}\odot P_{t_{M}}^{b}$, where
$\odot$ is a variant of the tensor product symbol $\otimes$, emphasizing
that the factors in the quantum history refer to different times,
and $P_{0^{-}}^{a}$is the projector on the energy eigenstate of $H_{L}$
with energy $a$ measured at time $t=0^{-}$, similarly for $P_{t_{M}}^{b}$.
And the corresponding chain operator $K\left(Y\right)$ is given by
the expression $K\left(Y\right)=P_{t_{M}}^{b}U\left(t_{M},0^{-}\right)P_{0^{-}}^{a}$
in the case of a quantum history $Y=P_{0^{-}}^{a}\odot P_{t_{M}}^{b}$
involving just two times, where $U\left(t_{M},0^{-}\right)$ is the
time evolution operator of the full Hamiltonian $H_{tot}$. It could
be easily verified that the joint probability distribution for the
quantum history $Y$ is
\begin{equation}
\Pr\left(Y\right)=\left\langle K^{\dagger}K\right\rangle \equiv\mathrm{Tr}\left\{ \rho^{ini}K^{\dagger}K\right\} ,
\end{equation}
where the superscript $\dagger$ stands for transpose conjugate and
from now on the angle brackets $\left\langle \ldots\right\rangle $
simply denotes ensemble average with respect to $\rho^{ini}$. Here
it is worth mentioning that this scheme for the joint probability
distribution could be readily extended to many-time measurement.

Taking the commutator $[\rho^{ini},\, P_{0^{-}}^{a}]=0$ into account,
we can verify that consistency conditions $\left\langle K\left(Y\right),K\left(Y'\right)\right\rangle _{\rho^{ini}}\equiv\mathrm{Tr}\left\{ \rho^{ini}K\left(Y\right)^{\dagger}K\left(Y'\right)\right\} =0$
for all $Y\neq Y'$ are fulfilled, which guarantees that we are working
with a consistent quantum framework. The consistent quantum framework,
combining the axiom of probability with the Born rule of the quantum
theory, is crucial to understand the probability aspect of the quantum
theory \cite{Griffiths2002}. Now we want to gain some insight into
the specific consistent quantum framework we study from thermal transport
point of view.

As we will consider later, the generating function (GF) for heat transfer
during the time $t_{M}$ is defined as
\begin{align}
\mathcal{Z}\left(\xi\right)= & \sum_{a,b}e^{i\xi\left(a-b\right)}\Pr\left(P_{0^{-}}^{a}\odot P_{t_{M}}^{b}\right),\label{eq: GF_def}
\end{align}
where the summation for $a$ and $b$ extends over all the eigenvalues
of $H_{L}$. Then based on this definition, we simply take the derivative
of $\mathcal{Z}\left(\xi\right)$ with respect to $i\xi$ and then
set $\xi=0$ to obtain the average heat transfer out of left lead
$L$
\begin{align}
 & Q_{L}\left(t_{M}\right)\nonumber \\
= & \sum_{a,b}\left(a-b\right)\Pr\left(P_{0^{-}}^{a}\odot P_{t_{M}}^{b}\right)\\
= & \mathrm{Tr}\left[\rho^{ini}H_{L}\right]-\mathrm{Tr}\left[\rho^{ini}U\left(0^{-},t_{M}\right)H_{L}U\left(t_{M},0^{-}\right)\right],
\end{align}
where properties $\left(P_{0^{-}}^{a}\right)^{2}=P_{0^{-}}^{a}$ and
$\sum_{a}P_{0^{-}}^{a}=1$ for the projector $P_{0^{-}}^{a}$, and
similarly for $P_{t_{M}}^{b}$ are used. Also we use the key relation
$[\rho^{ini},\, P_{0^{-}}^{a}]=0$, which assures that we are working
with a consistent quantum framework. Thus, we immediately realize
that
\begin{equation}
\frac{dQ_{L}\left(t_{M}\right)}{dt_{M}}=-\left\langle \frac{dH_{L}\left(t_{M}\right)}{dt_{M}}\right\rangle ,
\end{equation}
of which the right-hand side is just the natural definition of thermal
current $I_{L}$ out of the left lead at time $t=t_{M}$. Actually,
if the initial density matrix does not commute with $P_{0^{-}}^{a}$,
such as a steady-state density matrix, so that the framework we work
with is not consistent, then the thermal current deriving from GF
for the heat transfer will not be equal to the natural definition
of it.

\section{CUMULANT GENERATING FUNCTION\label{sec:CUMULANT-GENERATING-FUNCTION}}

As was mentioned before, we proceed to study the GF for the heat transfer
out of the left lead. According to the definition of the GF in Eq.~\eqref{eq: GF_def},
we can obtain
\begin{align}
\mathcal{Z}\left(\xi\right)= & \left\langle U_{\xi/2}\left(0^{-},t_{M}\right)U_{-\xi/2}\left(t_{M},0^{-}\right)\right\rangle ,
\end{align}
where $U_{-\xi/2}\left(t_{M},0^{-}\right)$ means evolution operator
associated with the modified total Hamiltonian $H_{tot}^{-\xi/2}\equiv e^{i\left(-\xi/2\right)H_{L}}H_{tot}e^{-i\left(-\xi/2\right)H_{L}}$,
and similarly for $U_{\xi/2}\left(0^{-},t_{M}\right)$. Transforming
to the interaction picture with respect to the free part of the modified
total Hamiltonian $H_{0}=H_{L}+H_{C}+H_{R}$, the GF for the heat
transfer becomes
\begin{align}
 & \mathcal{Z}\left(\xi\right)=\nonumber \\
 & \left\langle T_{\tau}e^{-\frac{i}{\hbar}\int_{C}d\tau\hat{u}_{L}\left(\hbar x_{\tau}+\tau\right)\left(V^{LC}\hat{u}_{C}\left(\tau\right)+V^{LR}\hat{u}_{R}\left(\tau\right)\right)+\hat{u}_{C}\left(\tau\right)V^{CR}\hat{u}_{R}\left(\tau\right)}\right\rangle ,\label{eq: GF_ini}
\end{align}
where $T_{\tau}$ is a $\tau$-ordering operator arranging operators
with earliest $\tau$ on the contour $C$ (from $0^{-}$ to $t_{M}$
and back to $0^{-}$) to the right, and a caret is put above operators
to denote their $\tau$ dependence with respect to the free Hamiltonian
such as $\hat{u}_{C}\left(\tau\right)=e^{\frac{i}{\hbar}H_{C}\tau}u_{C}e^{-\frac{i}{\hbar}H_{C}\tau}$,
and $x_{\tau}=-\xi/2$ with $\tau=t^{+}$ on the upper branch of the
contour $C$, while $x_{\tau}=+\xi/2$ with $\tau=t^{-}$ on the lower
branch.

The key step to evaluate GF is to rewrite the exponent in Eg.~\eqref{eq: GF_ini}
as $-\frac{i}{\hbar}\int_{C}d\tau_{1}d\tau_{2}\frac{1}{2}u^{T}\left(\tau_{1}\right)V\left(\tau_{1},\tau_{2}\right)u\left(\tau_{2}\right)$,
where

\begin{gather}
u^{T}\left(\tau\right)=\begin{bmatrix}\hat{u}_{L}\left(\hbar x_{\tau}+\tau\right) & \hat{u}_{C}\left(\tau\right) & \hat{u}_{R}\left(\tau\right)\end{bmatrix},\\
V\left(\tau_{1},\tau_{2}\right)=\begin{bmatrix}0 & V^{LC} & V^{LR}\\
V^{CL} & 0 & V^{CR}\\
V^{RL} & V^{RC} & 0
\end{bmatrix}\delta\left(\tau_{1},\tau_{2}\right),\\
\left(V^{T}\left(\tau_{1},\tau_{2}\right)=V\left(\tau_{1},\tau_{2}\right)\right).\nonumber
\end{gather}
 Here the generalized $\delta$-function $\delta\left(\tau_{1},\tau_{2}\right)$
is simply counterpart of the ordinary Dirac delta function on the
contour $C$, see, for example, Ref.~\cite{Wang2008}.

Then expanding the exponential to perform a perturbation expansion
and employing Feynman diagrammatic technique, especially Wick's theorem
and the linked cluster theorem, the CGF for the heat transfer can
be obtained to be
\begin{align}
\mathcal{\ln Z}\left(\xi\right)= & \frac{1}{2}\sum_{n=1}^{\infty}\mathrm{Tr}_{\left(j,\tau\right)}\left[\frac{1}{n}\left(Vg^{x}\right)^{n}\right]\\
= & \frac{1}{2}\sum_{n=1}^{\infty}\mathrm{Tr}_{\left(j,\tau\right)}\left[\frac{1}{n}\left(V_{red}g_{red}^{x}\right)^{n}\right]\\
= & -\frac{1}{2}\mathrm{Tr}\ln\left(I-\tilde{V}_{red}\tilde{g}_{red}^{x}\right),
\end{align}
where in the first equality
\begin{align}
 & g^{x}\left(\tau_{1},\tau_{2}\right)\nonumber \\
= & \begin{bmatrix}g^{L}\left(\hbar x_{\tau_{1}}+\tau_{1},\hbar x_{\tau_{2}}+\tau_{2}\right) & 0 & 0\\
0 & g^{C}\left(\tau_{1},\tau_{2}\right) & 0\\
0 & 0 & g^{R}\left(\tau_{1},\tau_{2}\right)
\end{bmatrix},
\end{align}
with the equilibrium contour-ordered Green functions
\begin{align}
g_{jk}^{\alpha}\left(\tau_{1},\tau_{2}\right)= & -\frac{i}{\hbar}\mathrm{Tr}\left\{ \frac{e^{-\beta_{\alpha}H_{\alpha}}}{\mathrm{Tr}\left(e^{-\beta_{\alpha}H_{\alpha}}\right)}T_{\tau}\left[\hat{u}_{j}^{\alpha}\left(\tau_{1}\right)\hat{u}_{k}^{\alpha}\left(\tau_{2}\right)\right]\right\} ,\\
 & \alpha=L,C,R\nonumber
\end{align}
and the notation $\mathrm{Tr}_{\left(j,\tau\right)}$ means trace
both in real space index $j$ and contour time $\tau$ such as $\mathrm{Tr}_{\left(j,\tau\right)}\left[Vg^{x}\right]=\int_{C}d\tau_{1}\int_{C}d\tau_{2}\mathrm{Tr}_{j}\left[V\left(\tau_{1},\tau_{2}\right)g^{x}\left(\tau_{2},\tau_{1}\right)\right]$;
in the second equality, considering the structure of the last expression,
instead of the full matrix we only need the finite reduced ones, which
make all the contribution to $\mathcal{\ln Z}\left(\xi\right)$, and
\begin{align}
V_{red}\left(\tau_{1},\tau_{2}\right)= & \delta\left(\tau_{1},\tau_{2}\right)\begin{bmatrix}0 & V_{red}^{LC} & V_{red}^{LR}\\
V_{red}^{CL} & 0 & V_{red}^{CR}\\
V_{red}^{RL} & V_{red}^{RC} & 0
\end{bmatrix}
\end{align}
is the generalized $\delta$-function $\delta\left(\tau_{1},\tau_{2}\right)$
times the reduced total coupling matrix obtained by deleting all the
zero column and row vectors of the full one except for the possible
zero vectors whose row or column indexes are the center (junction)
ones, and $g_{red}^{x}$ is the corresponding submatrix of $g^{x}$
just like $V_{red}$ of $V$; in the third equality a tilde above
matrix means discretized contour-time version of the corresponding
quantity such as $\left[\tilde{g}_{red}^{x}\right]_{\tau_{i}l,\tau_{j}n}=\left[g_{red}^{x}\right]_{ln}\left(\tau_{i},\tau_{j}\right)d\tau_{j}$
with an evenly spaced grid $\tau_{i}$ and $\tau_{j}$ along the contour
$C$ and $I$ is the identity matrix.

Introducing the Dyson equation
\begin{align}
 & G\left(\tau,\tau'\right)\nonumber \\
= & g\left(\tau,\tau'\right)+\int_{C}d\tau_{1}\int_{C}d\tau_{2}g\left(\tau,\tau_{1}\right)V\left(\tau_{1},\tau_{2}\right)G\left(\tau_{2},\tau'\right)\label{eq:Dyson1}\\
= & g\left(\tau,\tau'\right)+\int_{C}d\tau_{1}\int_{C}d\tau_{2}G\left(\tau,\tau_{1}\right)V\left(\tau_{1},\tau_{2}\right)g\left(\tau_{2},\tau'\right),\label{eq:Dyson2}
\end{align}
which actually defines $G=\begin{bmatrix}G^{LL} & G^{LC} & G^{LR}\\
G^{CL} & G^{CC} & G^{CR}\\
G^{RL} & G^{RC} & G^{RR}
\end{bmatrix}$ based on $g\left(\tau,\tau'\right)=g^{x\equiv0}\left(\tau,\tau'\right)$,
we easily realized that
\begin{align}
\tilde{G}_{red}= & \tilde{g}_{red}+\tilde{g}_{red}\tilde{V}_{red}\tilde{G}_{red}\label{eq:Dyson_red1}\\
= & \tilde{g}_{red}+\tilde{G}_{red}\tilde{V}_{red}\tilde{g}_{red}\label{eq:Dyson_red2}
\end{align}
still holds. And from now on, for notational simplicity, we omit all
the subscript $red$ with the understanding that these matrices are
of finite dimensions in the real space domain. Thus, employing Egs.~\eqref{eq:Dyson_red1}
and~\eqref{eq:Dyson_red2} along with the equality $\mathrm{Tr}\ln A=\ln\det A$,
we obtain

\begin{align}
 & \mathcal{\ln Z}\left(\xi\right)=-\frac{1}{2}\ln\det\left(I-d\right),\\
d\equiv & \begin{bmatrix}\tilde{V}^{LC} & \tilde{V}^{LR}\end{bmatrix}\begin{bmatrix}\tilde{G}^{CC} & \tilde{G}^{CR}\\
\tilde{G}^{RC} & \tilde{G}^{RR}
\end{bmatrix}\begin{bmatrix}\tilde{V}^{CL}\\
\tilde{V}^{RL}
\end{bmatrix}\tilde{g}_{L}^{A},\nonumber
\end{align}
where $\tilde{g}_{L}^{A}$ is the discretized contour-time version
of $g_{L}^{A}\left(\tau_{1},\tau_{2}\right)=g^{L}\left(\hbar x_{\tau_{1}}+\tau_{1},\hbar x_{\tau_{2}}+\tau_{2}\right)-g^{L}\left(\tau_{1},\tau_{2}\right)$.

If we introduce $\tilde{g}_{L}^{-1}$ satisfying $\tilde{g}_{L}^{-1}\tilde{g}_{L}=I$
and employ Eqs.~\eqref{eq:Dyson_red1} and~\eqref{eq:Dyson_red2},
we can simplify the CGF for heat transfer further to be
\begin{align}
\mathcal{\ln Z}\left(\xi\right)= & -\frac{1}{2}\ln\det\left(I-\tilde{g}_{L}^{-1}\left(\tilde{G}^{LL}-\tilde{g}_{L}\right)\tilde{g}_{L}^{-1}\tilde{g}_{L}^{A}\right),\label{eq:CGF_general}
\end{align}
which is valid in both transient and steady-state regimes.

\section{THE STEADY-STATE CGF\label{sec:THE-STEADY-STATE-CGF}}

Now, we proceed to evaluate the long-time limit of the CGF in Eg.~\eqref{eq:CGF_general}
called the steady-state CGF. Transforming Eg.~\eqref{eq:CGF_general}
from contour-time to real-time and then using the Keldysh rotation,
which is essentially an orthogonal transformation that
\begin{align}
\breve{A}\left(t_{1},t_{2}\right)= & O^{T}\sigma_{z}\begin{bmatrix}A\left(t_{1}^{+},t_{2}^{+}\right) & A\left(t_{1}^{+},t_{2}^{-}\right)\\
A\left(t_{1}^{-},t_{2}^{+}\right) & A\left(t_{1}^{-},t_{2}^{-}\right)
\end{bmatrix}O\\
\equiv & \begin{bmatrix}A^{r}\left(t_{1},t_{2}\right) & A^{K}\left(t_{1},t_{2}\right)\\
A^{\bar{K}}\left(t_{1},t_{2}\right) & A^{a}\left(t_{1},t_{2}\right)
\end{bmatrix}
\end{align}
with $O=\frac{1}{\sqrt{2}}\begin{bmatrix}1 & 1\\
-1 & 1
\end{bmatrix}$ and the Pauli $z$ matrix $\sigma_{z}=\begin{bmatrix}1 & 0\\
0 & -1
\end{bmatrix}$ appearing due to the transition from contour-time to real-time, see,
Eg. Ref.~\cite{Agarwalla2012}, we obtain
\begin{align}
\mathcal{\ln Z}\left(\xi\right)= & \frac{1}{2}\sum_{n=1}^{\infty}\mathrm{Tr}_{\left(j,t\right)}\left[\frac{1}{n}\left(\breve{g}_{L}^{-1}\left(\breve{G}^{LL}-\breve{g}_{L}\right)\breve{g}_{L}^{-1}\breve{g}_{L}^{A}\right)^{n}\right]\label{eq:CGF_keldsh}
\end{align}
where the notation $\mathrm{Tr}_{\left(j,t\right)}$ means trace both
in real space $j$ and real time $t$ such as $\mathrm{Tr}_{\left(j,t\right)}\left(\breve{A}\breve{B}\right)=\int_{0}^{t_{M}}dt_{1}\int_{0}^{t_{M}}dt_{2}\mathrm{Tr}_{j}\left[\breve{A}\left(t_{1},t_{2}\right)\breve{B}\left(t_{2},t_{1}\right)\right]$.

Before proceeding, a significant consideration comes that all kinds
of real-time versions of the contour-time Green function $G\left(\tau,\tau'\right)$
defined in Eqs.~\eqref{eq:Dyson1} or \eqref{eq:Dyson2} are not
necessarily time translationally invariant so that $\breve{G}^{LL}\left(t,t'\right)$
may not simply depend on the time difference $t-t'$. However, in
the long-time limit, i.e., $t_{M}\rightarrow\infty$, the time translationally
invariant part obtained from the lowest order of the Wigner transformation
will dominate the CGF \cite{Haug2008}. It is equivalent to saying
that $\breve{G}^{LL}\left(t,t'\right)=\breve{G}^{LL}\left(t-t'\right)$
is time translationally invariant in the long-time limit (higher order
terms of the product of the Wigner transformation have been ignored).

Consequently, setting $t_{M}\rightarrow\infty$, and Fourier transforming
Eq. \eqref{eq:CGF_keldsh}, we get
\begin{align}
 & \mathcal{\ln Z}\left(\xi\right)\nonumber \\
= & -\frac{1}{2}t_{M}\int_{-\infty}^{\infty}\frac{d\omega}{2\pi}\ln\det\left(I-\left(\breve{g}_{L}^{-1}\breve{G}^{LL}-I\right)\breve{g}_{L}^{-1}\breve{g}_{L}^{A}\right),\label{eq:CGF_steady}
\end{align}
where,
\begin{gather}
\breve{g}_{L}=\begin{bmatrix}g_{L}^{r} & g_{L}^{K}\\
0 & g_{L}^{a}
\end{bmatrix},\:\breve{G}^{LL}=\begin{bmatrix}G_{LL}^{r} & G_{LL}^{K}\\
0 & G_{LL}^{a}
\end{bmatrix},\\
\breve{g}_{L}^{A}=\frac{1}{2}\begin{bmatrix}a-b & a+b\\
-a-b & -a+b
\end{bmatrix},\\
a\equiv g_{L}^{>}\left(e^{-i\hbar\omega\xi}-1\right),\: b\equiv g_{L}^{<}\left(e^{i\hbar\omega\xi}-1\right)\nonumber
\end{gather}
are all in frequency space. For NEGF notations and some general relations
among Green\textquoteright{}s functions, we refer to Ref. \cite{Wang2008}.

To further simplify the steady-state CGF in Eq. $\eqref{eq:CGF_steady}$,
we use the formula $\det\begin{pmatrix}A & B\\
C & D
\end{pmatrix}=\det\left(AD-BC\right)$ in case of $\left[C,D\right]=0$ to reduce the dimension of the matrix
inside determinant by half. Therefore, the steady-state CGF is given
by
\begin{gather}
\mathcal{\ln Z}\left(\xi\right)=-t_{M}\int_{-\infty}^{\infty}\frac{d\omega}{4\pi}\ln\det \Big\{ I-\mathcal{T}_{\mathcal{G}}\left[\omega\right]\nonumber \\
\times\left[\left(e^{i\xi\hbar\omega}-1\right)f_{L}\left(1+f_{R}\right)+\left(e^{-i\xi\hbar\omega}-1\right)f_{R}\left(1+f_{L}\right)\right]\Big\} ,\label{eq:CGF_steady_final}
\end{gather}
where $\mathcal{T_{G}\left[\omega\right]\equiv}G_{LR}^{r}\tilde{\Gamma}_{R}G_{RL}^{a}\tilde{\Gamma}_{L}$
with
\begin{align}
\tilde{\Gamma}_{\left\{ L,R\right\} }\equiv & i\left[\left(g_{\left\{ L,R\right\} }^{a}\right)^{-1}-\left(g_{\left\{ L,R\right\} }^{r}\right)^{-1}\right]
\end{align}
 is the transmission matrix and $f_{\left\{ L,R\right\} }=\left\{ \exp\left(\beta_{\{L,R\}}\hbar\omega\right)-1\right\} ^{-1}$
is the Bose-Einstein distribution function for phonons.

In deriving it, we have used fluctuation dissipation theorem $e^{-\beta_{L}\hbar\omega}g_{L}^{>}\left[\omega\right]=g_{L}^{<}\left[\omega\right]=f_{L}\left(g_{L}^{r}-g_{L}^{a}\right)$
along with
\begin{align}
 & G_{LL}^{<}\nonumber \\
= & G_{LL}^{r}\left(-if_{L}\tilde{\Gamma}_{L}\right)G_{LL}^{a}+G_{LR}^{r}\left(-if_{R}\tilde{\Gamma}_{R}\right)G_{RL}^{a},\\
 & G_{LL}^{r}-G_{LL}^{a}-G_{LL}^{r}\left(g_{L}^{a-1}-g_{L}^{r-1}\right)G_{LL}^{a}\nonumber \\
= & G_{LR}^{r}\left(g_{R}^{a-1}-g_{R}^{r-1}\right)G_{RL}^{a},
\end{align}
due to $\left(g_{C}^{a}\right)^{-1}-\left(g_{C}^{r}\right)^{-1}=0$
and the Langreth theorem \cite{Haug2008} acting on Eqs.~\eqref{eq:Dyson_red1}
and \eqref{eq:Dyson_red2}. A computationally practical closed equation
for $G_{LR}^{r}$ could be found in Ref.~\cite{Li2012}. The formulas
Eq.~\eqref{eq:CGF_steady_final} and Eq.~\eqref{eq:CGF_general}
are our central results.

Now we recover the classical version of the CGF for the heat transfer in harmonic networks
without the lead-lead coupling, which was first derived in Ref.~\cite{Saito2011}
using the Langevin equation method. To this end we simply set the
lead-lead coupling $V^{LR}=0$ in Eqs.~\eqref{eq:Dyson_red1} and
\eqref{eq:Dyson_red2}, then use the Langreth theorem and the Fourier
transformation to obtain $G_{LR}^{r}=g_{L}^{r}V^{LC}G_{CR}^{r}=g_{L}^{r}V^{LC}G_{CC}^{r}V^{CR}g_{R}^{r}$
along with $G_{RL}^{a}=\left(G_{LR}^{r}\right)^{\dagger}$.

After setting $\hbar\rightarrow0$ and employing $G_{RL}^{a}\left[-\omega\right]=\left(G_{LR}^{r}\left[\omega\right]\right)^{T}$and
$\tilde{\Gamma}_{\left\{ L,R\right\} }\left[-\omega\right]=-\tilde{\Gamma}_{\left\{ L,R\right\} }\left[\omega\right]^{T}$,
we can get
\begin{align}
\lim_{t_{M}\rightarrow\infty}\frac{\mathcal{\ln Z}\left(\xi\right)}{t_{M}}= & -\int_{0}^{\infty}\frac{d\omega}{2\pi}\ln\det \Big\{ I-\mathcal{T}\left[\omega\right]\nonumber \\
 & \times k_{B}^{2}T_{L}T_{R}\left(i\xi\right)\left[i\xi+\left(\beta_{R}-\beta_{L}\right)\right]\Big\} ,
\end{align}

with
\begin{align}
\mathcal{T}\left[\omega\right]\equiv & G_{CC}^{r}\Gamma_{R}G_{CC}^{a}\Gamma_{L},\\
\Gamma_{\left\{ L,R\right\} }= & i\left[\Sigma_{\left\{ L,R\right\} }^{r}-\Sigma_{\left\{ L,R\right\} }^{a}\right]\\
\Sigma_{\alpha}^{\left\{ r,a\right\} }= & V^{C\alpha}g_{\alpha}^{\left\{ r,a\right\} }V^{\alpha C},\,\alpha=L,R.\nonumber
\end{align}

\section{THE SSFT AND CUMULANTS\label{sec:THE-SSFT-AND}}

According to the steady-state CGF in Eq.~\eqref{eq:CGF_steady_final},
one could easily verify that the GC fluctuation symmetry \cite{Gallavotti1995}
$\mathcal{Z}\left(\xi\right)=\mathcal{Z}\left(-\xi+i\left(\beta_{R}-\beta_{L}\right)\right)$
is still satisfied in this general set-up with lead-lead coupling.
And recall the definition of GF in Eq. \eqref{eq: GF_def}, we know
that the probability distribution for the heat transferred $Q_{L}$
is $\Pr\left(Q_{L}\right)=\frac{1}{2\pi\delta\left(0\right)}\int_{-\infty}^{\infty}d\xi \mathcal{Z}\left(\xi\right)e^{-i\xi Q_{L}}$.
Therefore, following the GC symmetry is the SSFT $\Pr\left(Q_{L}\right)=e^{\left(\beta_{R}-\beta_{L}\right)Q_{L}}\Pr\left(-Q_{L}\right)$,
which quantifies the violation of the second-law in the sense of probability.

Also, the CGF can be used to evaluate cumulants. Here we only focus
on steady-state cumulants of heat transfer. As illustrated in Sec.
\ref{sec:MODEL-AND-CONSISTENT}, the steady current is closed related
to the first cumulant so that
\begin{align}
I_{L}^{ss}= & \lim_{t_{M}\rightarrow\infty}\frac{d}{dt_{M}}\left(\frac{\partial\mathcal{\ln Z}\left(\xi\right)}{\partial\left(i\xi\right)}\mid_{\xi=0}\right)\nonumber \\
= & \int_{0}^{\infty}\frac{d\omega}{2\pi}\hbar\omega\left(f_{L}-f_{R}\right)\mathrm{Tr}\mathcal{T}_{\mathcal{G}}\left[\omega\right],
\end{align}
where, $\mathcal{Z}\left(0\right)=1$ is used. This generalized Caroli
formula with lead-lead coupling was given very recently in Ref. \cite{Li2012}
based on the definition of current directly, which gave us some valuable
hints on the form of the steady-state CGF. The second cumulant describing
the fluctuation of the heat transferred is obtained by taking the
second derivative of steady-state CGF with respect to $i\xi$ and
then setting $\xi=0$, which is
\begin{align}
\left\langle \left\langle Q_{L}^{2}\right\rangle \right\rangle =\, & t_{M}\int_{-\infty}^{\infty}\frac{d\omega}{4\pi}\left(\hbar\omega\right)^{2}\Big\{ \left(f_{L}+f_{R}+2f_{L}f_{R}\right)\mathrm{Tr}\mathcal{T}_{\mathcal{G}}\nonumber \\
 & +\left(f_{L}-f_{R}\right)^{2}\mathrm{Tr}\mathcal{T}_{\mathcal{G}}^{2}\Big\} .
\end{align}
Higher-order cumulants are also systematically given by corresponding
higher-order derivatives.

After some experiences on first few order cumulants of heat transfer,
we want to discuss the trick suggested by Di Ventra mentioned in Sec.~\ref{sec:INTRODUCTION} that repartitioning the total Hamiltonian
to avoid the inevitable coupling between leads in real nanoscale or
mesoscopic system when calculating steady current. Now whether this
trick is applicable to the evaluation of higher-order cumulant (fluctuation)
of the heat transfer in steady state boils down to checking whether $\mathrm{Tr}\mathcal{T}_{G,old}^{n}=\mathrm{Tr}\mathcal{T}_{G,new}^{n}$
hold for all $n$ less than or equal to the corresponding order of
the cumulant one wants, where $\mathcal{T}_{G,old}\left(\mathcal{T}_{G,new}\right)$
is the transmission matrix before (after) repartitioning the total
Hamiltonian. Though giving a general verification is difficult, $\mathrm{Tr}\mathcal{T}_{G,old}^{n}=\mathrm{Tr}\mathcal{T}_{G,new}^{n},$$\,\forall n$
is indeed true in a one-dimensional central ring model, in which there
is only one particle in the center junction connected with two semi-infinite
spring chain leads and the interaction between the two nearest particles
in the two leads respectively exists (in this case, both $\mathcal{T}_{G,old}$
and $\mathcal{T}_{G,new}$ are just a number). One step forward, if
one think of CGF of heat transfer as the complete knowledge of the
steady state, we can claim that the steady state is partition-independent
after verification of $\mathrm{Tr}\mathcal{T}_{G,old}^{n}=\mathrm{Tr}\mathcal{T}_{G,new}^{n},\,\forall n$
or equivalently $\mathcal{\ln Z}_{old}\left(\xi\right)=\mathcal{\ln Z}_{new}\left(\xi\right)$
in Eq. \eqref{eq:CGF_steady_final}. Then we can partly answer a question
raised by Caroli $\mathit{et\: al.}$ regarding the (non)equivalence
between the partitioned and partition-free approaches \cite{Caroli1971},
which recently was partly settled by explicitly constructing a non-equilibrium
steady state through adiabatically turning on an electrical bias between
the leads \cite{Cornean2012}.

\section{SUMMARY}

We examine the statistics of heat transfer during time $t_{M}$ in
a general lead-junction-lead quantum system, in which coupling between
leads has been taken into account. To this end, a consistent quantum
framework was introduced to derive the CGF valid in both transient
and long-time regimes using the NEGF method. Also, the implication
of consistency of the quantum framework was briefly discussed from
thermal transport point of view. After that, a compact form of the
steady-state CGF was obtained, following which the GC symmetry and
the SSFT was verified. In addition, first few cumulants were given
and generalized Caroli formula was recovered. Furthermore, some valuable
hints with respect to the rigorous proof for whether fluctuation of
heat transfer in steady state is partition-independent have been offered.
\begin{acknowledgments}
We thank Juzar Thingna and Lifa Zhang for insightful discussions.
This work is supported in part by URC Grant No. R-144-000-257-112.
\end{acknowledgments}

\bibliographystyle{apsrev4-1}
\bibliography{C_Generating_F}

\end{document}